\documentstyle[12pt,psfig]{article}
\newcounter{one}
\newcounter{two}
\begin{document}
\newcommand{\kptwo}{KP~\Roman{two}}
\newcommand{\kpe}{Kadomtsev--Petviashvili equation}
\newcommand{\kpetwo}{Kadomtsev--Petviashvili~\Roman{two} equation}
\newcommand{\beq}{\begin{equation}}
\newcommand{\eeq}{\end{equation}}
\newcommand{\sgn}{\mathop{\rm sgn}\nolimits}
\newcommand{\const}{\mathop{\rm const}\nolimits}
\newcommand{\psif}[1]{\Psi(\lambda_{#1},x,y)}
\newcommand{\chif}[1]{\chi(\lambda_{#1},x,y)}
\newcommand{\cpsif}[1]{\Psi^*(\lambda_{#1},x,y)}
\newcommand{\cchif}[1]{\chi^*(\lambda_{#1},x,y)}
\newcommand{\chifo}[2]{\chi_{#2}(\lambda_{#1},x,y)}
\newcommand{\cchifo}[2]{\chi^*_{#2}(\lambda_{#1},x,y)}
\newcommand{\tchifo}[2]{\tilde\chi_{#2}(\lambda_{#1},x,y)}
\newcommand{\lr}{\lambda_R}
\newcommand{\li}{\lambda_I}
\newcommand{\Lo}[1]{L_0(\lambda_{#1})}
\newcommand{\cLo}[1]{L_0^*(\lambda_{#1})}
\newcommand{\tGl}[1]{\hat{\tilde G}(\lambda_{#1})}
\newcommand{\Gl}[1]{\hat G(\lambda_{#1})}
\newcommand{\cGl}[1]{\hat G^*(\lambda_{#1})}
\newcommand{\phik}[1]{\phi_{#1}(\lambda,x,y)}
\newcommand{\cphik}[1]{\phi_{#1}^*(\lambda,x,y)}
\newcommand{\dphik}[1]{\check\phi_{#1}(\lambda,x,y)}
\newcommand{\cdphik}[1]{\check\phi_{#1}^*(\lambda,x,y)}

\newtheorem{thm}{Theorem}
\newtheorem{lem}{Lemma}
\newtheorem{rem}{Remark}
\setcounter{section}{0}
\input epsf.sty
\input amssym.def
\input amssym.tex

\begin{center}
{\Large Nonsingularity of the direct scattering transform for the 
\kptwo \ equation with real exponentially decaying at infinity potential.}

\medskip

{\bf P.G.Grinevich \footnote{This paper was partially fulfilled during the
visit of the author to Lecce (Italy) in June-July 1995. This visit was a part
of the INTASS cooperation program grant No 93-166. The author was also 
supported by Soros International Scientific Foundation grant MD~8000 and 
by the Russian Foundation for Fundamental Studies grant 95-01-00755.}

\smallskip

Landau Institute for Theoretical Physics, Kosygina 2, Moscow, 117940, Russia,

e-mail: pgg@landau.ac.ru}
\end{center}

\medskip

{\bf Abstract.} {\it  We study the direct spectral transform for the 
heat equation, associated with the \kpetwo. We show, that for real 
nonsingular exponentially decaying at infinity potentials the direct
problem is nonsingular for arbitrary large potentials. Earlier this
statement was proved only for potentials, satisfying the ``small norm''
assumption.}

{\bf \arabic{section} Introduction.} The \kpe \ (KP)
$$
\left(u_t-6uu_x+u_{xxx}\right)_x=-2\alpha^2u_{yy}
$$
was historically the first equation with 2 spatial variables integrated
by the inverse scattering transform (IST) method (Druma, Zakharov-Shabat) 
\cite{D}, \cite{Z-Sh}. The auxiliary linear operator, associated 
with KP reads as
\beq
L=\alpha\partial_y-\partial^2_x+u(x,y).
\label{Lop}
\eeq

Usually $\alpha^2$ is assumed to be real. Without loss of generality
we can assume that $\alpha^2=\pm 1$ so it is sufficient to consider the 
following two cases:
\begin{enumerate}
\item $\alpha=i$ and we have KP~\Roman{one} or unstable KP.
\item $\alpha=1$ and we have \kptwo \ or stable KP.
\end{enumerate}

These two cases are essentially different from the analytical point of 
view. The scattering transform for KP~\Roman{one} with a decaying at
infinity potential was constructed by Manakov \cite{Man} in terms of
the nonlocal Riemann problem. A more regular method for constructing
such transform was suggested in \cite{A-F}. The scattering transform for
the \kptwo \ equation with a decaying at infinity potential was 
constructed by Ablowitz, Bar Yaakov and Fokas \cite{A-BY-F} in terms
of the so-called $\bar\partial$ problem. (Additional information and
the technical details can be found in review paper \cite{F-S}).

\begin{rem} In our paper we do not plan to discuss the periodic
theory of KP, developed by Krichever (see \cite{Kr1}, \cite{Kr2}). But
we would like to point out that in the periodic theory the difference 
between KP~\Roman{one} and
\kptwo \ is also very essential. For example in \cite{Kr2} it was shown that
the direct transform is well--defined for \kptwo \ and any periodic
potential can be approximated by a finite-gap one, but for KP\Roman{one}
these problems are still open. 
\end{rem}
\begin{rem} A new approach to the decaying at infinity theory, based
on the resolvent technique was suggested in recent paper by Boiti,
Pempinelli, Pogrebkov and Polivanov \cite{B-P-P-P}. This approach can be 
extended to more general classes of potentials including for example,
the one-dimensional solitons.
\end{rem}

The main technical tool used in the spectral theory of operators with
decaying at infinity potentials is the Fredholm theory of integral
equations. Consider the following equation
\beq
\left(\hat 1 +\hat A \right) x = b
\label{Feq}
\eeq
where $\hat A$ is an integral (compact) operator, $b$ is a given vector 
in the Hilbert space and $x$ is an unknown vector. If the operator norm
of $\hat A$ is less then 1 equation (\ref{Feq}) is uniquely solvable 
$$
x=\left(\hat 1-\hat A+\hat A^2 -\hat A^3 +\ldots  \right) b,
$$
but if the norm of $\hat A$ is greater then 1 the situation is more
complicated (see for example \cite{Fan}).

Equation (\ref{Feq}) has at least one solution if and only if the 
right-hand side $b$ is orthogonal to the kernel of the adjoint equation
\beq
<b,x_k^*> =0 \ \hbox{for all} \ x_k^* \ \hbox{such that} 
\left(\hat 1 +\hat A^* \right) x_k^* = 0.
\label{Falt}
\eeq
This property is called the Fredholm alternative.

In the scattering theory the norm of the integral operator can be 
estimated via the norm of the potential in the direct problem or
via the norm of the scattering data in the inverse problem. 
Hence if these norms are sufficiently small the unique solvability of the
integral equations of the scattering theory can be easily proved
and we essentially simplify the situation. This case was studied 
absolutely strictly (see \cite{F-S}).

In general if the small norm assumption is not fulfilled, then the solutions
of the integral equations have singularities for some special values
of the spectral parameter and we have to introduce additional spectral
data corresponding to these singular points. But fortunately there are
some important examples such that the unique solvability of the integral
equations can be proved without the ``small norm'' assumption.

One of such examples was found by S.P.Novikov and the author \cite{G-N}.
Consider the inverse problem for the two-dimensional Schr\"odinger 
operator at a fixed negative energy or the inverse problem for the operator 
(\ref{Lop}) with $\alpha=1$. The wave function is defined as a solution
of the $\bar\partial$ equation. Assume that only the real Schr\"odinger
operators without first order terms or only the heat operators with real
potentials are considered. Then the scattering data satisfies some 
additional (rather simple) necessary and sufficient conditions. 
If these conditions are fulfilled, then 
the unique solvability of the $\bar\partial$ equation 
for arbitrary large data follows from the
theory of the generalized analytic functions.
Another important example is the solvability of the inverse problem 
for (\ref{Lop}) with $\alpha=i$ and real $u(x,y)$ proved 
by Zhou \cite{Zh}.

In our paper we show that the direct scattering transform for the 
heat operator 
$$
L=\partial_y-\partial^2_x+u(x,y).
$$
with a real exponentially decaying
at infinity potential is nonsingular without the ``small norm'' 
assumption. (This spectral problem corresponds to the 
\kptwo \ equation).
It is very likely that the exponential decay rate is not too essential 
by it is not clear how to weaken this condition in our proof.

The plan of our paper is the following. In the first section we recall
the scheme of the scattering transform for the \kptwo \ equation. In the
second section we assume that equations of the direct scattering
have nonzero homogeneous solutions and study the properties of these
solutions. We show that these solutions have to satisfy some orthogonality
conditions. In the third section we show that from these orthogonality
conditions it follows that these homogeneous solutions generate no
singularities in the scattering transform.

The author is grateful to the Department of Physics, University of
Lecce for the hospitality.
The author is also grateful to L.Bogdanov, M.Boiti, B.Konopelchenko and
A.Pogrebkov for stimulating discussions.

\refstepcounter{section}
{\bf \arabic{section} The scattering transform for the \kptwo \ equation.}
\label{section1}
In our paper we study the heat operator
\beq
L=\partial_y-\partial^2_x+u(x,y).
\label{Hop}
\eeq
with a real smooth potential $u(x,y)$. Denote by $r$ the distance from
the origin $r=\sqrt{x^2+y^2}$. We shall put on $u(x,y)$ one of the following 
assumptions at infinity:
\begin{itemize}
\item[D1)] $u(x,y)$ decays at infinity faster than $r^{-2}$, i.e., 
there exist a constant $\varepsilon >0$ and a collection of constants
$c_{mn}$ such that 
$$
\left |  \frac{\partial^{m+n}u(x,y)}{\partial x^m \partial y^n}\right | <
\frac{c_{mn}}{(1+r)^{2+\epsilon}} \ \hbox{for all} \ m,n \ge 0.
$$
\item[D2)] $u(x,y)$ belongs to the Schwartz class, i.e., 
there exist a constant $\varepsilon >0$ and a collection of constants
$c^k_{mn}$ such that 
$$
\left |  \frac{\partial^{m+n}u(x,y)}{\partial x^m \partial y^n} \right | <
\frac{c^k_{mn}}{(1+r)^{2+k}} \ \hbox{for all} \ m,n,k \ge 0.
$$
\item[D3)] $u(x,y)$ exponentially decays at infinity, i.e., 
there exist a constant $\varepsilon >0$ and a collection of constants
$c_{mn}$ such that 
$$
\left |  \frac{\partial^{m+n}u(x,y)}{\partial x^m \partial y^n} \right | <
c_{mn}e^{-\epsilon r} \ \hbox{for all} \ m,n \ge 0.
$$
\end{itemize}

The direct scattering transform suggested in \cite{A-BY-F} (for more
details see \cite{F-S}) is the following: 

Let $\psif{}$ be a solution of the heat equation
\beq
L\Psi(\lambda,x,y) = 0
\label{Heq}
\eeq
and 
\beq
\psif{}=e^{\lambda x+\lambda^2 y} \chif{},\ \chif{}=1+o(1) \ \hbox{as} \ 
x^2+y^2 \rightarrow \infty,
\label{as}
\eeq
where $\lambda$ is a complex parameter. In our text we use the following
agreement: the notation $f(\lambda)$, $\lambda \in \Bbb C$ does 
{\underline {not}} mean that $f$ is holomorphyc in $\lambda$, i.e.,
we do {\underline {not}} assume that $\bar\partial f(\lambda)=0$.

It is convenient to introduce new variables:
\begin{eqnarray}
\lr=\hbox{Re}\lambda,\ \li=\hbox{Im}\lambda,\ z=x+2\lambda y,\ 
\bar z=x+2\bar\lambda y, \nonumber\\ 
\partial_z=\frac1{4i\li}\left( \partial_y-
2\bar\lambda\partial_x \right ),\ \partial_{\bar z}=\frac{-1}{4i\li }\left ( 
\partial_y-2\lambda\partial_x \right ).  
\label{zvar}
\end{eqnarray}
Let
$$
\Lo{}=\partial_y-\partial_x^2-2\lambda\partial_x=-4i\li\partial_{\bar z}-
\left (\partial_z+\partial_{\bar z}\right)^2.
$$
The function $\chif{}$ satisfies the following equation:
\beq
\Lo{}\chif{}+u(x,y)\chif{}=0.
\label{Chieq}
\eeq
Let $G(\lambda,x,y)$ be the Green function of the operator $\Lo{}$
$$
\Lo{} G(\lambda,x,y)=\delta(x)\delta(y).
$$
The function $G(\lambda,x,y)$ reads as
\beq
G(\lambda,x,y)=\frac{1}{4\pi^2} \int\int \frac{e^{i(px+qy)}dpdq}
{p^2+iq-2i\lambda p}.
\label{Greenf}
\eeq
Equation (\ref{Chieq}) with the boundary condition (\ref{as}) is 
equivalent to the following integral equation
\beq
\chif{}=1-\Gl{} u(x,y)\chif{}
\label{Inteq}
\eeq
where $\Gl{}$ denotes the integral operator
$$
\Gl{} f(\lambda,x,y)=\int\int G(\lambda,x-x',y-y')f(\lambda,x',y')dx'dy'.
$$
The function $G(\lambda,x,y)$ is not holomorphyc in $\lambda$ but
$$
\frac{G(\lambda,x,y)}{\partial\bar\lambda}=-\frac{i \sgn\li}{2\pi}
e^{-2i\li x -4i \li\lr y}.
$$
Differentiating (\ref{Inteq}) w.r.t. $\bar\lambda$ we get
\beq
\frac{\chif{}}{\partial\bar\lambda}=-\Gl{} u(x,y) 
\frac{\chif{}}{\partial\bar\lambda} +T(\lambda)e^{-2i\li x -4i \li\lr y},
\label{barder}
\eeq
where
\beq
T(\lambda)=\frac{i \sgn\li}{2\pi}b(\lambda), \ 
b(\lambda)=\int\int u(x,y)\chif{} e^{2i\li x + 4i \li\lr y}dxdy.
\label{dbardata}
\eeq
{}From (\ref{barder}) it follows that
$$
\frac{\chif{}}{\partial\bar\lambda}=T(\lambda)\chi_1(\lambda,x,y),
$$
where $\chi_1(\lambda,x,y)$ is defined by the following integral equation
$$
\chi_1(\lambda,x,y)=-\Gl{} u(x,y) \chi_1(\lambda,x,y)+ 
e^{-2i\li x -4i \li\lr y}.
$$
{}From the symmetry properties of the Green function $G(\lambda,x,y)$
it follows that
$$
\chi_1(\lambda,x,y)=e^{-2i\li x -4i \li\lr y}\chi(\bar\lambda,x,y).
$$
Finally we get
$$
\frac{\chif{}}{\partial\bar\lambda}=
T(\lambda)e^{-2i\li x -4i \li\lr y}\chi(\bar\lambda,x,y)
$$
or, equivalently
$$
\frac{\psif{}}{\partial\bar\lambda}=T(\lambda)\Psi(\bar\lambda,x,y)
$$
The function $b(\lambda)$ is the scattering data for the \kptwo \ equation. 
The potential $u(x,y)$ is real if and only if
\beq
b(\bar\lambda)=\bar b(\lambda).
\label{real}
\eeq
All calculations from this section are correct if the integral equation
(\ref{Inteq}) has an unique nonsingular solution for any $\lambda \in \Bbb C$.
The solvability of (\ref{Inteq}) is discussed below.

We need also the adjoint equation to (\ref{Heq})
$$
\left[ -\partial_y-\partial_x^2+u(x,y) \right ] \cpsif{}=0
$$
where 
$$
\cpsif{}=e^{-\lambda x-\lambda^2 y} \cchif{}, \cchif{}=1+o(1) \ \hbox{as} \ 
x^2+y^2 \rightarrow \infty.
$$
Let
$$
\cLo{}=-\partial_y-\partial_x^2+2\lambda\partial_x=4i\li\partial_{\bar z}-
\left (\partial_z+\partial_{\bar z}\right)^2.
$$
The function $\cchif{}$ satisfies the following equation:
$$
\cLo{}\cchif{}+u(x,y)\cchif{}=0.
$$
The adjoint Green function reads as
$$
G^*(\lambda,x,y)=\frac{1}{4\pi^2} \int\int \frac{e^{i(px+qy)}dpdq}
{p^2-iq+2i\lambda p}.
$$
The function $\cchif{}$ satisfies the following integral equation
\beq
\cchif{}=1-\cGl{} u(x,y)\cchif{}
\label{cInteq}
\eeq

\refstepcounter{section}
{\bf \arabic{section} \label{section2}
Homogeneous solutions of the direct scattering 
integral equation.} If the homogeneous part of (\ref{Inteq}) 
\beq
\chif{}=-\Gl{} u(x,y)\chif{}
\label{hInteq}
\eeq
has no
nonzero solutions for any $\lambda$ then the direct scattering transform
is well defined (see for example \cite{F-S}). Assume now that equation
(\ref{hInteq}) has nonzero solutions for some $\lambda=\lambda_0$,
$u(x,y)$ is a real smooth function, decaying at infinity faster than $r^{-2}$
(D1 decay condition from the section~\ref{section1}). In this section we
study the properties of such solutions.

If $\li=0$ then (\ref{Inteq}) is a Volterra type equation, i.e.,
$$
G(\lambda,x,y) =0 \ \hbox{for all}\ y<0
$$
and (\ref{cInteq}) has no nonzero solutions. Hence without loss of generality
we may assume that $\li\ne0$.

It is convenient to introduce a new function
$$
\tilde\chif{}=e^{i\li x +2i \li\lr y}\chif{}.
$$
The corresponding Green function reads as
$$
\tilde G(\lambda,x,y)=\frac{1}{4\pi^2} \int\int \frac{e^{i(px+qy)}dpdq}
{p^2-\li^2+i(q-2\lr p)}.
$$
The function $\tilde\chif{}$ satisfy the following integral equation
$$
\tilde\chif{}= e^{i\li x +2i \li\lr y}-\tGl{} u(x,y)\tilde\chif{}.
$$
The homogeneous equation reads as
\beq
\tilde\chif{}=-\tGl{} u(x,y)\tilde\chif{}
\label{htildeq}
\eeq
The homogeneous equation(\ref{htildeq}) is real and we have the obvious
one-to-one correspondence between the solutions of (\ref{hInteq})
and (\ref{htildeq}). From the Fredholm theory it follows that the spaces 
of these solutions are finite-dimensional. 
Let $\tchifo{}{1}$, \ldots, $\tchifo{}{k}$
be a real basis of solutions of (\ref{htildeq}), 
$\chifo{}{1}$, \ldots, $\chifo{}{k}$ be the corresponding
solutions of (\ref{hInteq}). 

\begin{lem}
\label{lemma1}
The Green function $\tilde G(\lambda,x,y)$ has the following asymptotic 
expansion as $x^2+y^2\rightarrow\infty$:
\begin{eqnarray}
\tilde G(\lambda,x,y)= 
e^{i\li x +2i \li\lr y}\left\{1 + \sum_{k=1}^{\infty}
\left[\frac{i}{4\li}\left ( \partial_{\bar z}+2 \partial_z+
\partial_{\bar z}^{-1} \partial_z^2  \right)  \right]^k\right\}
\frac{i\sgn\li}{2\pi z}+\nonumber\\
+\hbox{\rm Complex conjugate terms} \ 
\label{Gexp}
\end{eqnarray}
All the monomials in this expansion have the form $c_{mn}\bar z^m/z^n$
$n>2m\ge 0$.Hence the operator $\partial_{\bar z}^{-1} \partial_z^2$
is well defined
$$
\partial_{\bar z}^{-1} \partial_z^2 \frac{\bar z^m}{z^n} =
\frac{n(n+1)}{m+1}\frac{\bar z^{m+1}}{z^{n+2}}.
$$
The proof of this Lemma is rather standard.
\end{lem}
Consider the following collection of eigenfunctions of $\Lo{}$:
\begin{eqnarray}
\phik{k}=\left.e^{-\mu x -\mu^2 y} \partial^k_{\mu} e^{\mu x+\mu^2y}
\right|_{\mu=\lambda},
\dphik{k}=\left.e^{-\mu x -\mu^2 y} \partial^k_{\bar\mu} 
e^{\bar\mu x+\bar\mu^2y} \right|_{\mu=\lambda}, \nonumber\\
\Lo{}\phik{k}=\Lo{}\dphik{k}=0,
\nonumber
\end{eqnarray}
and the adjoint collection
\begin{eqnarray}
\cphik{k}=\phi_k(\lambda,-x,-y),\ 
\cdphik{k}=\check\phi_k(\lambda,-x,-y), 
\nonumber\\
\cLo{}\cphik{k}=\cLo{}\cdphik{k}=0,
\label{phiadj}
\end{eqnarray}
\begin{lem}
\label{lemma2} 
\Roman{one}) Let $\tchifo{}0$ be a real nonzero solution of the homogeneous
integral equation (\ref{htildeq}), $\chifo{}0=e^{-i\li x -2i \li\lr y}
\tchifo{}0$ be the corresponding solution of (\ref{hInteq}). Then there are 
two possibilities:
\begin{enumerate}
\item Either $\tchifo{}0$, $\chifo{}0$ are Schwartz functions in $(x,y)$, 
i.e., they decay at infinity faster than any degree of $z$.
\item Or 
$$
\tchifo{}0=
e^{i\li x +2i \li\lr y}\frac{c}{z^n}+
e^{-i\li x -2i \li\lr y}\frac{\bar c}{\bar z^n}
+ o\left(\frac{1}{z^{n}} \right),
$$
$$
\chifo{}0=
\frac{c}{z^n}+
e^{-2i\li x -4i \li\lr y}\frac{\bar c}{\bar z^n}
+ o\left(\frac{1}{z^{n}} \right).
$$
\end{enumerate}
\Roman{two}) If $\chifo{}0=O\left(\frac{1}{z^{n+2}} \right)$ then
$$
\int\int \cphik{k}u(x,y)\chifo{}0dxdy =
$$
\beq
=\int\int \cdphik{k}u(x,y)\chifo{}0dxdy =0 \ \hbox{for}\ k=0,\ldots,n.
\label{Ort}
\eeq
\end{lem}
{\bf Proof of Lemma~\ref{lemma2}.} To get the asymptotic expansion 
for $\tchifo{}0$ as $r\rightarrow\infty$ we substitute (\ref{Gexp}) 
in the right hand side of (\ref{htildeq}). Then we transform all 
monomials in the expansion of
$G(\lambda,x-x',y-y')$ to the following form
$$
\frac{(\bar z-\bar z')^m}{(z-z')^n}=\frac{\bar z^m}{z^n}\cdot 
\left(1-\frac{\bar z'}{\bar z}\right)^m\cdot
\left(1+\frac{z'}{z}+\left(\frac{z'}{z}\right)^2+\left(\frac{z'}{z}\right)^3
+\ldots \right)^n.
$$
Integrating the right-hand side of (\ref{htildeq}) in $x'$, $y'$ we 
get some formal expansion of $\tchifo{}0$ and the coefficients of this 
expansion read as
$$
\int\int P_{mn}(z,\bar z) u(x,y) \tchifo{}0 dx dy,
$$
where $P_{mn}$ are some polynomials in $z$, $\bar z$. Of course some
of these integrals may diverge. 

First consider the terms of this expansion decaying as $1/z$, then decaying as
$1/z^2$, then as $1/z^3$ and so on. In each order of decay rate we have only
finite number of monomials. It is rather easy to check that the integrals 
representing the lowest order nonzero coefficients in this expansion 
converge and these terms give us the right asymptotics of $\tchifo{}0$.
If all coefficients are equal to zero all integral converges and the
function $\tchifo{}0$ decays faster than any degree of $r$.

Assume now that at least one of the terms in the asymptotic expansion
is nonzero. By definition (we use now that $u(x,y)=o(1/z^2)$ as $z\rightarrow
\infty$) we have
$$
\Lo{}\chifo{}0=-4i\li\partial_{\bar z}-
\left (\partial_z+\partial_{\bar z}\right)^2\chifo{}0=
o\left(\frac1{z^2}\right)\chifo{}0.
$$
Thus the first nonvanishing coefficient in the asymptotics for $\chifo{}0$
contains no $\bar z$. This completes the proof of the first part of 
Lemma~\ref{lemma2}.

Let $\kappa(\lambda,x,y)$ be one of the functions $\cphik{k}$, $\cdphik{k}$
$0\le k\le n$. By definition
$$
I=\int\int\kappa(\lambda,x,y)u(x,y)\chifo{}0dxdy=
$$
$$
=-\int\int\kappa(\lambda,x,y)\Lo{} \chifo{}0dxdy.
$$
Integrating by parts we get
$$
I=\int\int\chifo{}0\cLo{}\kappa(\lambda,x,y)dxdy+\hbox{Boundary terms}.
$$
If $\chifo{}0=O\left(\frac{1}{z^{n+2}} \right)$ the boundary terms
vanishes. Applying (\ref{phiadj}) we get the orthogonality property $I=0$.
Lemma~\ref{lemma2} is proved.
\begin{lem}
\label{lemma3}
Let $\Psi(x,y)$ be a global real solution of the heat equation 
\beq
\left( \partial_y-\partial^2_x+u(x,y) \right) \Psi(x,y)=0
\label{Heq2}
\eeq
($u(x,y)$ is assumed to be a real smooth function defined in the whole plane).

Denote by $\cal D$ the open set of points $(x,y)$ such that $\Psi(x,y)\ne0$.
Denote by ${\cal D}_j$ the arcwise connected components of $\cal D$.

Assume that the component ${\cal D}_j$ is locally bounded in the 
$x$-direction, i.e., for any $y_1$, $y_2$ the intersection of the 
set ${\cal D}_j$ with the strip $y_1<y<y_2$ is bounded. 

Then the set ${\cal D}_j$ is unbounded from below in the $y$-direction,
i.e., there exists a path $[0,\infty)\rightarrow{\cal D}_j$, 
$t\rightarrow(x(t),y(t))$ such that
$$
\frac{\partial y(t)}{\partial t} <0,\ y(t)\rightarrow-\infty\ \hbox{as}\ 
t\rightarrow+\infty.
$$
\end{lem}
\begin{rem}
In Lemma~\ref{lemma3} the functions $\Psi(x,y)$ and $u(x,y)$ are assumed
to be real, smooth and defined for all $(x,y)$, but they may grow arbitrary
fast as $r\rightarrow\infty$.
\end{rem}
{\bf Proof of Lemma~\ref{lemma3}.} Consider the following function
$$
J(y)=\int\limits_{{\cal L}(y)} \frac{1}{2}\Psi^2(x,y) dx
$$
where ${\cal L}(y)$ denotes the intersection of ${\cal D}_j$ with 
the line $y=\const$.
The function $\Psi^2(x,y)$ vanishes on the boundary of ${\cal D}_j$. Hence
$$
\frac{\partial J(y)}{\partial y}=\int\limits_{{\cal L}(y)}\Psi(x,y)
\partial_y\Psi(x,y)dx
$$
(the terms arousing from differentiation by the position of the
boundaries of ${\cal L}(y)$ are equal to 0). From (\ref{Heq2}) it follows 
that
$$
\frac{\partial J(y)}{\partial y}=-\int\limits_
{{\cal L}(y)}u(x,y)\Psi^2(x,y)dx+
\int\limits_{{\cal L}(y)}\Psi(x,y)\partial^2_x\Psi(x,y)dx.
$$
The function $\Psi(x,y)$ vanishes on the boundary of ${\cal D}_j$. Hence 
integrating by parts we get
$$
\frac{\partial J(y)}{\partial y}=-\int\limits_{{\cal L}(y)}u(x,y)\Psi^2(x,y)dx-
\int\limits_{{\cal L}(y)}(\partial_x\Psi(x,y))^2dx<2J(y)\max_{x\in{\cal L}(y)}
\left|u(x,y)\right|.
$$
For any $y<y_0$ such that $J(y_0)>0$ we have the following estimate
$$
J(y)>e^{c_{y,y_0}(y-y_0)}J(y_0)>0,\ \hbox{where}\ 
c_{y,y_0}=2\max_{x\in{\cal L}(\tilde y),y\le \tilde y\le y_0} 
\left|u(x,\tilde y)\right|.
$$
Thus if ${\cal L}(y_0)$ is not empty then ${\cal L}(y)$ is not empty
for any $y<y_0$. This completes the proof.

Now we are ready to prove:
\begin{thm}
\label{theorem1}
Let $\chifo{}0$ be a real nonzero solution of the homogeneous
integral equation (\ref{hInteq}) where $u(x,y)$ is a real smooth function
decaying at infinity faster then $r^{-2}$ (D1 decay condition). Then 
\begin{enumerate}
\item $\chifo{}0$ decays at infinity faster than any degree of 
$r$, i.e., belongs to the Schwartz class. 
\item  The function $u(x,y)\chifo{}0$ is orthogonal to the functions
$\cphik{k}$, $\cdphik{k}$ for all $k\ge 0$
$$
\int\int \cphik{k}u(x,y)\chifo{}0dxdy =
$$
\beq
=\int\int \cdphik{k}u(x,y)\chifo{}0dxdy =0.
\label{Ort2}
\eeq
\end{enumerate}
\end{thm}
{\bf Proof of Theorem \ref{theorem1}.} Without loss of generality we can
assume that the functions 
$$\tchifo{}0=e^{i\li x+2i \li\lr y}\chifo{}0,\ 
\Psi_0(\lambda,x,y)=e^{\lambda x+\lambda^2 y}\chifo{}0
$$ 
are real. The zeroes of $\tchifo{}0$ coincide with the zeroes of 
$\Psi_0(\lambda,x,y)$, $\Psi_0(\lambda,x,y)$ is a global solution
of (\ref{Heq2}). Thus the zeroes of $\tchifo{}0$ have the properties
formulated in Lemma~\ref{lemma3}.

By Lemma~\ref{lemma2} if the function $\tchifo{}0$ is not from the 
Schwartz class then the leading term of $\tchifo{}0$ reads as
\beq
\tchifo{}0\sim
e^{i\li x +2i \li\lr y}\frac{c}{z^n}+
e^{-i\li x -2i \li\lr y}\frac{\bar c}{\bar z^n}.
\label{chias}
\eeq
It is easy to check that (\ref{chias}) approximate $\tchifo{}0$ with 
the first derivative. The zeroes of the leading term are non-degenerate
hence they approximate the zeroes of $\tchifo{}0$ for sufficiently large $z$.
Let $C$ be a sufficiently big circle.

\begin{center}
\mbox{\epsfysize=2.in
\epsffile{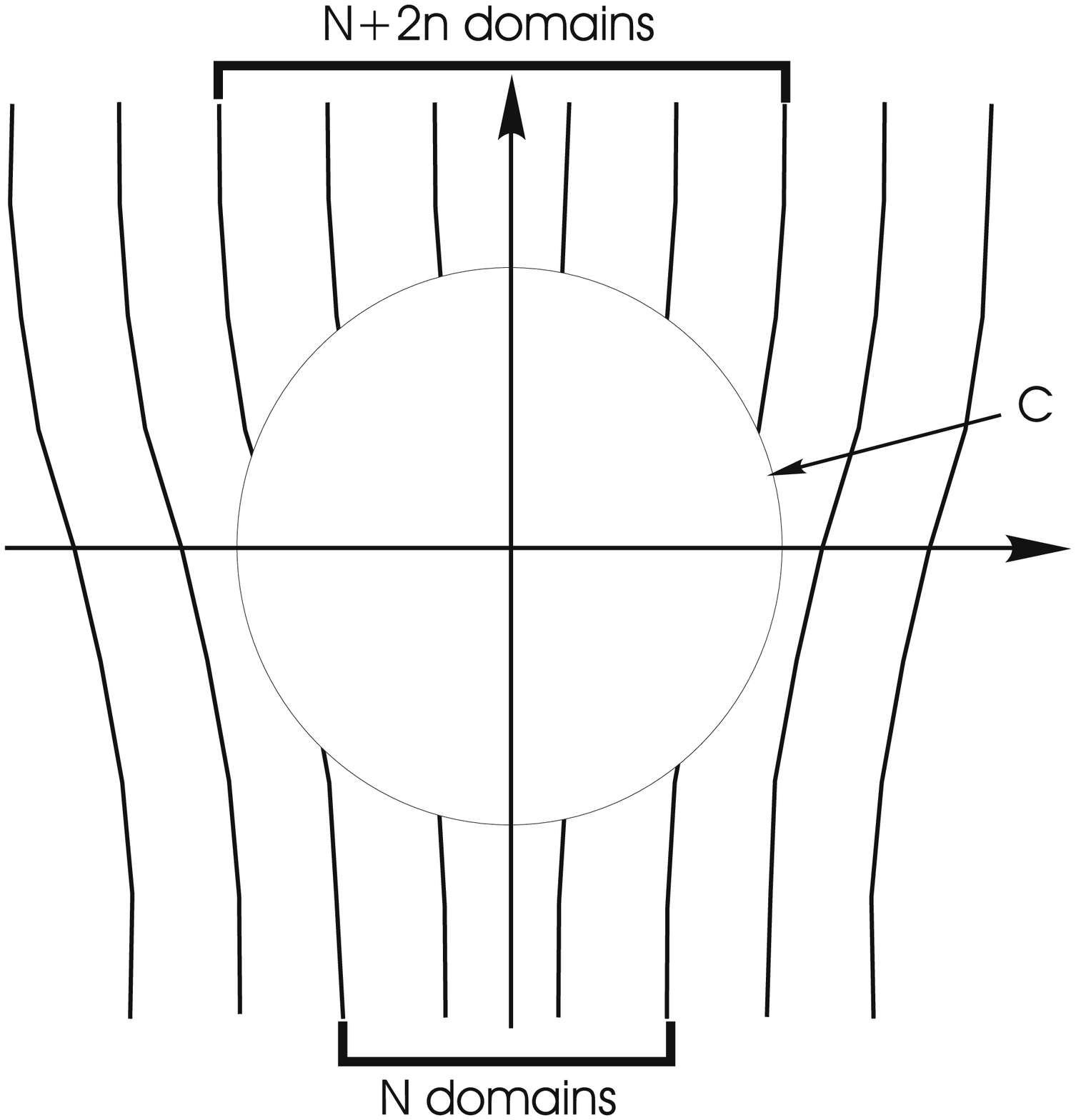}}

Fig 1.
\end{center}

Let us draw the zeroes of $\tchifo{}0$ outside $C$ . Consider the set of 
domains ${\cal D}_j$ intersecting the circle $C$ (the domains ${\cal D}_j$
were defined in Lemma~\ref{lemma3}). Let $N$ be the number of such domains 
coming from the direction $y=-\infty$. Elementary analysis of (\ref{chias}) 
shows that the number of domains coming from $y=+\infty$ and 
intersecting $C$ is equal to $N+2n$. Hence at least one of the domains 
coming from $y=+\infty$ vanishes as $y\rightarrow-\infty$. But by 
Lemma~\ref{lemma3} it is impossible. Thus assuming that $\chifo{}0$ has 
nonvanishing terms in the asymptotic expansion we get a contradiction.
Comparing this result with Lemma~\ref{lemma2} we complete the proof.

\refstepcounter{section}
{\bf \arabic{section} \label{section3} Regularity of the direct spectral
transform.} Assume now that $u(x,y)$ is a real smooth potential decaying
at infinity faster then any degree of $r$ (D2 decay condition). 

Let us call a point $\lambda_0$ regular if for $\lambda=\lambda_0$ 
the homogeneous integral equation (\ref{hInteq}) has no nonzero solutions. 
Otherwise the point $\lambda_0$ will be called irregular.
The aim of this section is to study the properties of the wave function
$\psif{}$ at irregular points.

In section \ref{section2} we defined the functions $\phik{k}$, 
$\dphik{k}$ for $k\ge0$. Let us define the following formal series
$$
\phik{k}=\left\{1 + \sum_{k=1}^{\infty}
\left[\frac{i}{4\li}\left ( \partial_{\bar z}+2 \partial_z+
\partial_{\bar z}^{-1} \partial_z^2  \right)  \right]^k\right\}z^k,
$$
$$
\dphik{k}=e^{-2i\li x-4i \li\lr y}\bar\phik{k}, \ 
$$
\beq
\cphik{k}=\phi_k(\lambda,-x,-y),\ 
\cdphik{k}=\check\phi_k(\lambda,-x,-y).
\label{formphi}
\eeq
For $k\ge0$ these series contain only finite number of nonzero terms and
coincide with (\ref{phiadj}).

\begin{thm}
\label{theorem2}
Let $u(x,y)$ be a real smooth potential, decaying at infinity faster then any
degree of $r$, $a_0$,\ldots,$a_n$, $b_0$,\ldots,$b_n$ be arbitrary
constants. Then for any $\lambda$ the following integral equation
\beq
\chif{}=\sum_{k=0}^n\left( a_k\phik{k}+b_k\dphik{k}  \right)-
\Gl{} u(x,y)\chif{}
\label{Inteq2}
\eeq
has at least one nonsingular solution with the following asymptotics as
$z\rightarrow\infty$:
$$
\chif{}=\sum_{k=-\infty}^n\left( a_k\phik{k}+b_k\dphik{k}  \right).
$$
The coefficients $a_k$, $b_k$ with $k<0$ are uniquely defined by the
coefficents $a_k$, $b_k$ with $k\ge0$.
\end{thm}
{\bf Proof of Theorem \ref{theorem2}.} If $\lambda$ is a regular point
then this statement is absolutely standard (see for example \cite{F-S}).
If $\lambda$ is an irregular point then (\ref{Inteq2}) has at least one 
nonsingular solution if and only if the non-homogeneous part is
orthogonal to the kernel of the adjoint equation (see for example \cite{Fan}).
By definition the adjoint equation to the integral equation (\ref{hInteq}) 
reads as
\beq
f(\lambda,x,y)=-u(x,y)\tGl{}f(\lambda,x,y)
\label{adjeq}
\eeq
(we use here $(AB)^*=B^*A^*$). The basis of solutions of 
(\ref{adjeq}) is formed by the functions $u(x,y)\cchifo{}{1}$, \ldots, 
$u(x,y)\cchifo{}{k}$ where $\cchifo{}{1}$, \ldots, $\cchifo{}{k}$ 
is some basis of solutions of the homogeneous part of (\ref{cInteq})
$$
\cchif{}=-\cGl{} u(x,y)\cchif{}.
$$
Thus we have the following orthogonality conditions:
$$
\int\int \phik{k}u(x,y)\cchifo{}{k}dxdy =
$$
\beq
=\int\int \dphik{k}u(x,y)\cchifo{}{k}dxdy =0.
\label{aOrt}
\eeq
Conditions (\ref{aOrt}) are adjoint to the orthogonality conditions
(\ref{Ort}) from Theorem~\ref{theorem1} and can be proved absolutely in
the same way.

The function $u(x,y)$ vanishes at infinity faster then any degree of $r$. 
Then asymptotically for any $N>0$ we have
\beq
\Lo{}\chif{}=o(z^{-N}). 
\label{asymptzero}
\eeq
Using the same approach as Lemma~\ref{lemma2} we can construct asymptotic 
expansion for $\chif{}$. Substituting this expansion in (\ref{asymptzero}) 
we obtain that the asymptotics is formed by linear combinations of $\phik{k}$, 
$\dphik{k}$. 

By Theorem~\ref{theorem1} if $\lambda$ is an irregular point then all
homogeneous solutions decay at infinity faster then any degree of
$z$; hence they do not affect the asymptotic expansion.
Theorem~\ref{theorem2} is proved.
\begin{rem}
Theorem \ref{theorem2} says that for any $\lambda$ we have at least one
solution of (\ref{Inteq2}). But we have not proved that these solutions
are continuous in $\lambda$. This question needs a more detailed analysis.
\end{rem}

Let $\lambda_0$ be an irregular point. By Theorem~\ref{theorem2} the 
direct scattering equation (\ref{Heq}) has at least one solution with
the asymptotics (\ref{as}) at the point $\lambda_0$. Our next step is 
to construct a formal solution of (\ref{Heq}) in the neighborhood
of $\lambda_0$. Consider the following formal series in
$(\lambda-\lambda_0)$, $(\bar\lambda-\bar\lambda_0)$
\beq
\psif{}=e^{\lambda_0 x+\lambda_0^2 y} \sum_{m,n\ge0}
\frac{(\lambda-\lambda_0)^m(\bar\lambda-\bar\lambda_0)^n}{m!n!}
\psi_{mn}(x,y),
\label{formsol}
\eeq
where the functions $\psi_{mn}(x,y)$ are defined by the following
properties:
\begin{enumerate}
\item All $\psi_{mn}(x,y)$ satisfy (\ref{Chieq}) with $\lambda=\lambda_0$.
\item $\psi_{mn}(x,y)$ are smooth in $x$, $y$ and 
\beq
\psi_{mn}(x,y)=\sum_{k=-\infty}^{\max(m,n)}c_k^{mn} \phi_k(\lambda_0,x,y)+
\sum_{k=-\infty}^{\max(m,n)}d_k^{mn} \check\phi_k(\lambda_0,x,y)
\label{psimnexp}
\eeq
as $z\rightarrow\infty$.
\item In all orders of perturbation theory in $(\lambda-\lambda_0)$, 
$(\bar\lambda-\bar\lambda_0)$ the asymptotic condition (\ref{as}) is
fulfilled.
\end{enumerate}

The last property needs some explanations. Substituting (\ref{psimnexp})
to (\ref{formsol}) we get
$$
\Psi(\lambda,x,y)=e^{\lambda_0 x+\lambda_0^2 y}\Phi_1(\lambda,x,y)+
e^{\bar\lambda_0 x+\bar\lambda_0^2 y}\Phi_2(\lambda,x,y)
$$
where
$$
\Phi_1(\lambda,x,y)=
\sum_{m,n\ge0}\sum_{k=-\infty}^{\max(m,n)} c_k^{mn}
\frac{(\lambda-\lambda_0)^m(\bar\lambda-\bar\lambda_0)^n}{m!n!}
\phi_k(\lambda_0,x,y)
$$
$$
\Phi_2(\lambda,x,y)=
\sum_{m,n\ge0}\sum_{k=-\infty}^{\max(m,n)} d_k^{mn}
\frac{(\lambda-\lambda_0)^m(\bar\lambda-\bar\lambda_0)^n}{m!n!}
\bar\phi_k(\lambda_0,x,y)
$$
Let 
$$
\Xi(\lambda,x,y)=e^{(\lambda_0-\lambda) x+(\lambda_0^2-\lambda^2) y}
\Phi_1(\lambda,x,y),
$$
$$ 
\check\Xi(\lambda,x,y)=e^{(\bar\lambda_0-\bar\lambda) x+
(\bar\lambda_0^2-\bar\lambda^2) y}\Phi_2(\lambda,x,y)
$$
Expanding the exponents in $(\lambda-\lambda_0)$, $(\bar\lambda-\bar\lambda_0)$
we get some formal series
$$
\Xi(\lambda,x,y)=
\sum_{m,n\ge0}
\frac{(\lambda-\lambda_0)^m(\bar\lambda-\bar\lambda_0)^n}{m!n!}
\xi_{mn}(x,y)
$$
$$
\check\Xi(\lambda,x,y)=
\sum_{m,n\ge0}
\frac{(\lambda-\lambda_0)^m(\bar\lambda-\bar\lambda_0)^n}{m!n!}
\check\xi_{mn}(x,y)
$$

The functions $\xi_{mn}(x,y)$, $\check\xi_{mn}(x,y)$ are some asymptotic 
Laurent series in $z$, $\bar z$. The property 3) means the following:
\beq
\xi_{mn}(x,y)=\delta_{m0}\delta_{n0}+o(1), \check\xi_{mn}(x,y)=o(1) \ 
\hbox{as}\ z\rightarrow\infty.
\label{xicond}
\eeq

\begin{thm}
\label{theorem3}
Let $u(x,y)$ be a real smooth potential, decaying at infinity faster then any
degree of $r$, $\lambda=\lambda_0$ be an irregular point. Then 
\begin{enumerate}
\item The equation of direct scattering (\ref{Heq}) has at least one 
formal solution (\ref{formsol}) with the properties 1)--3). 
\item The functions $\psi_{mn}(x,y)$ are defined uniquely up to adding
arbitrary solutions of the homogeneous equation (\ref{hInteq}). 
\item The constants $c_k^{mn}$, $d_k^{mn}$ are uniquely determined by
the properties 1)--3).
\item The ``scattering data'' $b(\lambda)$ is uniquely defined as a 
formal Taylor series in $(\lambda-\lambda_0)$, $(\bar\lambda-\bar\lambda_0)$.
\end{enumerate}
\end{thm}
{\bf Proof of Theorem~\ref{theorem3}.} 
Suppose $c_k^{mn}$, $d_k^{mn}$ are arbitrary constants such that 
$c_k^{mn}=d_k^{mn}=0$ for $k>\max(m,n)$. Then the formal series 
$\Xi(\lambda,x,y)$, $\check\Xi(\lambda,x,y)$ are well--defined 
in all orders of $(\lambda-\lambda_0)$, $(\bar\lambda-\bar\lambda_0)$ and 
satisfy
\beq
[\Lo{0} -2(\lambda-\lambda_0)\partial_x] \Xi(\lambda,x,y)=0, \ 
[L_0(\bar\lambda_0) -2(\bar\lambda-\bar\lambda_0)\partial_x] 
\check\Xi(\lambda,x,y)=0.
\label{Xieq}
\eeq
By Theorem~\ref{theorem2} for any $m$, $n$ and arbitrary $c_k^{mn}$, 
$d_k^{mn}$ $0\le k\le\max(m,n)$ we have at least one solution of 
(\ref{Inteq2}). This solutions is defined up to adding arbitrary solutions
of the homogeneous equation (\ref{hInteq}), the constants $c_k^{mn}$, 
$d_k^{mn}$ with $k<0$ are uniquely defined. Hence to prove the theorem
it is sufficient to show the existence of an unique collection of 
$c_k^{mn}$, $d_k^{mn}$, $k\ge 0$ such that the property 3) is fulfilled.

We calculate these constants by induction. Putting $\lambda=\lambda_0$ we 
get $c_0^{00}=1$, $d_0^{00}=0$. Assume that for some $l$ we know all 
coefficients $c_k^{mn}$, $d_k^{mn}$ with $m+n<l$. Consider the function
$\xi^{mn}(x,y)$, $m+n=l$. By (\ref{Xieq}) 
$$
\Lo{0}\xi^{mn}(x,y) -2m\partial_x\xi^{m-1\, n}(x,y)=0.
$$
By (\ref{xicond}) 
$$
\partial_x\xi^{m-1\, n}(x,y)=o(1) \ \hbox{as}\ z\rightarrow\infty .
$$
Hence 
$$
\Lo{0}\xi^{mn}(x,y)=o(1), \ 
\xi^{mn}(x,y)=\sum_{k=0}^{\max(m,n)}\upsilon_k^{mn}
\phi_k(\lambda_0,x,y)+o(1).
$$
{}From the definition of $\Xi(\lambda,x,y)$ it follows that
$$
\xi^{mn}(x,y)=F(\lambda,x,y,c_k^{mn},d_k^{mn})+ 
\sum_{k=0}^{\max(m,n)}c_k^{mn}\phi_k(\lambda_0,x,y)+o(1),
$$
where $F(\lambda,x,y,c_k^{mn},d_k^{mn})$ is a linear function of 
$c_k^{mn}$, $d_k^{mn}$ with $m+n<l$.

Thus there exists an unique collection of $c_k^{mn}$, $0\le k\le\max(m,n)$
such that $\xi^{mn}(x,y)=o(1)$. Similar analysis of 
$\check\Xi(\lambda,x,y)$ gives us the coefficients $d_k^{mn}$, $m+n$=l.

To define $b(\lambda)$ consider the asymptotic expansion of $\psif{}$
for large $x$, $y$
$$
\psif{}=e^{\lambda x+\lambda^2 y}\left[1+\frac{c_{-1}}{z}+\ldots\right]+
e^{\bar\lambda x+\bar\lambda^2 y}\left[\frac{d_{-1}}{\bar z}+\ldots\right].
$$
The function $b(\lambda)$ is defined by (\ref{dbardata}). Integrating
(\ref{dbardata}) by parts we get
$$
T(\lambda)=d_{-1}(\lambda), b(\lambda)=2\pi i\sgn{\li} d_{-1}(\lambda).
$$
At the previous step of the proof we have defined the function 
$d_{-1}(\lambda)$ as a formal series in $(\lambda-\lambda_0)$,
$(\bar\lambda-\bar\lambda_0)$. Thus $b(\lambda)$ is uniquely defined
as a formal series. This completes the proof.

Unfortunately the results of Theorem~\ref{theorem3} are not sufficient
to prove that the scattering data $b(\lambda)$ is a nonsingular function
because the series of perturbation theory may diverge. Assume now that
the potential $u(x,y)$ decay at infinity exponentially (D3 condition).
Then the scattering data $b(\lambda)$ is a ratio of two analytic 
functions of two real variables $\lr$, $\li$. For the Schr\"odinger 
operator the proof of analyticity can be found in \cite{RN}, the scheme 
of this proof can also be applied to the heat operator without 
serious modifications. Thus if $u(x,y)$ exponentially decays at 
infinity then the series of perturbation theory for $b(\lambda)$
converges and from Theorem~\ref{theorem3} it follows that the
functions $b(\lambda)$ and $\psif{}$ are regular in $\lambda$.
Let us recall that for nonsingular spectral data with the 
reality constraint (\ref{real}) the corresponding potential and the
wave function $\psif{}$ are uniquely defined by $\b(\lambda)$ without
the small norm assumption.
Hence we have proved the main result of the paper:

\begin{thm}
\label{theorem4}
Let $u(x,y)$ be a real smooth potential, exponentially decaying at infinity 
(D3 condition), $L$ be the heat operator (\ref{Hop}) associated with the 
\kptwo \ equation. Then the direct spectral transform for $L$ 
described in section~\ref{section1} is nonsingular for arbitrary 
large $u(x,y)$.
\end{thm}

\begin{rem}
The scattering transform for the heat operator can be obtained by an appropriate
limiting procedure from the scattering transform for the two-dimensional
Schr\"odinger operator at a fixed negative energy studied in \cite{G-N}.
These two scattering problems look rather similar. But now is is clear
that these two problems are essentially different at least in one aspect.
In the theory of the Schr\"odinger operator with real decaying at 
infinity potential if our fixed energy level
is located above the ground state then we always have singularities in
the spectral transform. It is not clear now how to pose the inverse
scattering problem for the singular scattering data 
(some preliminary results can be found in \cite{G-N}).
Thus in some sense the scattering transform for the heat operator 
is simpler (at least for sufficiently fast decaying potentials).
\end{rem}

\end{document}